\documentclass{llncs}
\usepackage{amsmath}
\usepackage{amsfonts}
\usepackage{amssymb}
\usepackage{graphicx}
\usepackage{hyperref}
\begin{document}
\title{Reasoning Under Threat: Symbolic and Neural Techniques for Cybersecurity Verification}
\author{Sarah Veronica}
\institute{}

\maketitle

\begin{abstract}
Cybersecurity demands rigorous and scalable techniques to ensure system correctness, robustness, and resilience against evolving threats. Automated reasoning, encompassing formal logic, theorem proving, model checking, and symbolic analysis, provides a foundational framework for verifying security properties across diverse domains such as access control, protocol design, vulnerability detection, and adversarial modeling. This survey presents a comprehensive overview of the role of automated reasoning in cybersecurity, analyzing how logical systems --- including temporal, deontic, and epistemic logics—are employed to formalize and verify security guarantees. We examine state-of-the-art tools and frameworks, explore integrations with AI for neural-symbolic reasoning, and highlight critical research gaps, particularly in scalability, compositionality, and multi-layered security modeling. The paper concludes with a set of well-grounded future research directions, aiming to foster the development of secure systems through formal, automated, and explainable reasoning techniques.
\end{abstract}

\section{Introduction}
Cybersecurity has become a paramount concern in our increasingly interconnected world, with diverse application domains such as healthcare, supply chain management, and the Internet of Things~\cite{atzori2010internet} facing sophisticated and evolving threats. The complexity and security-critical nature of modern cyber systems necessitate rigorous analysis and verification to ensure their correctness, reliability, and resilience against potential vulnerabilities. Logic and automated reasoning play a pivotal role in providing formal guarantees for these crucial properties of cybersecurity. By employing various logical frameworks and automated techniques, researchers and practitioners can formally model, specify, and verify different aspects of cyber systems, ranging from the intricate logic of security protocols to the correctness of access control mechanisms and the robustness against attacks. This survey aims to provide a comprehensive overview of the current state of research at the intersection of logic and automated reasoning for cybersecurity. It delves into the diverse types of logic utilized, the automated reasoning techniques employed, the specific applications addressed, the limitations encountered, and the promising directions for future research in this rapidly evolving field.

This paper addresses the fundamental challenge of ensuring the reliability and security of cyber systems through the application of formal methods. To this end, it seeks to answer the following research questions:
\begin{enumerate}
\item What logical systems (e.g., temporal, deontic, epistemic, higher-order logic) have been applied to model and verify critical security properties such as confidentiality, integrity, authentication, and access control?
\item How are automated reasoning techniques—including theorem proving, model checking, and SMT solving—used to analyze and formally verify the correctness and robustness of security protocols and systems?
\item What is the current state of research in using formal methods and automated reasoning for vulnerability detection and analysis, including static and dynamic techniques across software, hardware, and network domains?
\item What are the emerging research directions at the intersection of AI and formal reasoning for cybersecurity, and what technical foundations are needed to support scalable, compositional, and explainable security verification?
\end{enumerate}

This survey advances the state of research at the nexus of formal logic, automated reasoning, and cybersecurity by offering a comprehensive and technically grounded synthesis of existing literature. Through an in-depth analysis of seminal works and recent advancements, the paper unpacks how logical systems and reasoning tools have been harnessed to formalize, analyze, and secure complex cyber infrastructures. Beyond mapping current capabilities, it surfaces critical blind spots in scalability, tool integration, and applicability to emerging domains. The survey further charts a forward trajectory by outlining technically rigorous and strategically relevant research directions, including hybrid AI-logical methods and compositional security verification, grounded in formal and mathematical principles. In doing so, it provides both a snapshot of the field's current maturity and a blueprint for its evolution.

\section{A Unified View of Logical Frameworks and Reasoning Tools in Cybersecurity}

Formal logic and automated reasoning serve as the cornerstones of rigorous cybersecurity analysis. Logic provides the formal languages needed to specify security properties such as confidentiality, authentication, and integrity, while automated reasoning techniques enable the systematic verification of these properties at scale. Different logical systems offer varying expressive power and suitability for particular classes of security challenges.

First-order logic (FOL) has long served as a foundation for axiomatic modeling of security properties and system behavior. Temporal logics, including Linear Temporal Logic (LTL), are widely used for specifying and verifying time-dependent security properties, such as protocol state transitions and replay attack resilience. Deontic logic, which expresses permissions, obligations, and prohibitions, is especially relevant for access control policies, compliance verification, and security governance. More expressive frameworks, such as higher-order logic (HOL), enable formalization of complex system architectures and layered security models.

These logical systems are realized through practical formal methods and reasoning tools that automate verification tasks. The K framework~\cite{rosu2010overview}, Lean-based Clear, and Coq-based ConCert enable rigorous specification and formal verification of security protocols and software modules. Advanced efforts such as those by Cai et al. demonstrate how refinement calculus and separation logic can be enhanced with automated tools for scalable, sound verification. These efforts build on a rich tradition of foundational verification frameworks.

Automated reasoning also plays a key role in discharge of proof obligations and satisfiability checks. SMT solvers such as Z3~\cite{de2008z3} are tightly integrated into security verification pipelines in tools like Lean and K. Recent work explores the use of SMT-based optimization techniques in static analysis for vulnerability detection and protocol hardening.

Theorem proving and static analysis have been employed to verify both low-level code and high-level security policies. Applications of formal reasoning span from graph-based encrypted traffic analysis to secure communication protocol verification using vector modulation logics. Cyber-physical and federated systems—including IoT architectures—pose unique security challenges that have been addressed through logic-based verification frameworks emphasizing explainability and privacy.

The interaction between AI and formal logic opens new avenues for specification inference, vulnerability detection, and proof automation. Hybrid neural-symbolic approaches are gaining traction for security applications, including recent efforts that explore LLM-guided proof synthesis and formal explanation generation.

Security in distributed systems, particularly secure communication protocols and consensus mechanisms, has also benefited from formal specification and model checking. Tools such as PRISM have been used to evaluate probabilistic security guarantees under partial synchrony or Byzantine assumptions. Formal analysis of incentive compatibility, game-theoretic adversarial strategies, and trust assumptions is exemplified by the CheckMate framework and related work.

Finally, the application of formal verification in industry—such as in high-assurance blockchain platforms like Cardano and Tezos—demonstrates the growing importance of logic-based methodologies for real-world secure systems. Mi-Cho-Coq's verification of the Michelson language and IOHK’s Haskell-based research reflect a broader trend toward adopting formal methods as first-class security engineering tools.

\section{Formal Methods for Secure Protocols and Vulnerability Analysis}

Formal methods and automated reasoning provide a rigorous foundation for modeling, verifying, and analyzing security-critical systems. This section explores the role of logical frameworks, verification environments, and reasoning techniques in three core cybersecurity applications: secure protocol verification, formal design and analysis of communication protocols, and automated vulnerability detection across software and hardware systems.

\subsection{Verified Security Protocols: From Specification to Implementation}

Security protocols must guarantee properties such as confidentiality, integrity, and authentication even under adversarial conditions. Formal verification plays a pivotal role in ensuring these guarantees through precise, logic-based specifications and automated proof frameworks. Environments such as Coq (via ConCert), Lean (via Clear), and the K framework offer powerful platforms for protocol verification. These tools allow protocols to be encoded as logical assertions, supporting the verification of key properties through theorem proving and symbolic automation.

The use of refinement types has further increased the expressiveness and safety of formal specifications. Security-focused verification techniques often incorporate SMT solvers and model checkers to validate complex properties such as key secrecy, mutual authentication, and tamper resistance. Approaches that combine symbolic execution, static analysis, and logic-based verification help reduce false positives and negatives, improving assurance and automation.

Scaling these techniques to real-world protocols requires modularity and compositionality. The literature demonstrates frameworks for compositional verification of large-scale systems. Translation from verified specifications to executable code is a critical step in bridging theory and deployment, supported by tools such as those in. Verified implementations have been deployed in sectors including finance and healthcare, as shown in, and verification has extended to cover real-time and resource-constrained environments such as embedded and IoT systems.

\subsection{Logical Modeling and Analysis of Secure Communication Protocols}

Designing secure communication protocols requires formal models that accurately represent state transitions, adversarial capabilities, and probabilistic behavior. Logical methods—including state machines, process algebras, temporal logic, and epistemic models—enable rigorous design and analysis of such systems. Probabilistic frameworks like Continuous Time Markov Chains (CTMCs) support the modeling of stochastic security behavior and uncertain message delivery, as demonstrated in the analysis of randomized protocols.

Temporal logics such as LTL and CTL are particularly effective for expressing critical properties like safety and liveness in secure communication scenarios. Model checking techniques have been applied extensively to analyze Byzantine Fault-Tolerant (BFT) protocols and ensure robustness in the presence of faulty or malicious participants. Probabilistic model checking tools such as PRISM and PRISM+ have been used to explore network-level properties and failure scenarios.

More expressive models such as threshold automata and temporal epistemic logic enable reasoning about distributed trust, partial knowledge, and cryptographic guarantees. As secure communication protocols grow in complexity, compositional reasoning becomes vital. Logic-based frameworks that support asynchronous communication and hybrid systems allow modeling of precise timing and ordering constraints in secure messaging.

The growing prominence of federated and cross-domain systems introduces new challenges for security reasoning. Federated learning and multi-party protocols must coordinate trust, privacy, and communication under incomplete knowledge. Research on privacy-preserving coordination and federated trust models is increasingly important. Game-theoretic reasoning, combined with SAT/SMT-based symbolic utilities, has been applied to analyze strategic behavior and incentive structures in adversarial communication settings.

\subsection{Automated Reasoning for Vulnerability Detection and Threat Analysis}

Identifying vulnerabilities in complex systems requires systematic analysis of control flow, data dependencies, and threat surfaces. Automated reasoning techniques—especially those built on symbolic logic—enable scalable vulnerability detection in software, hardware, and network infrastructures. Tools like Mythril and Oyente apply symbolic execution to enumerate execution paths and detect vulnerabilities such as reentrancy, overflow, and unauthorized access. These systems use SMT solvers and constraint encodings to reason about safety violations under adversarial inputs.

Recent approaches expand these capabilities through modular engines, theorem proving integration, and semantic reasoning backends, improving precision and coverage. Static analysis methods based on logical inference have been used to verify security properties such as non-interference and secure information flow. Graph-based reasoning over encrypted communication flows has proven effective for network vulnerability assessment, while formal logics like vector modulation have enabled analysis of low-level communication channels.

Threat modeling, a key step in proactive security engineering, is often grounded in logic-based policy languages and epistemic frameworks. These approaches have been used to formally specify access control, analyze resilience to denial-of-service attacks, and synthesize defense strategies through logic programming. Integration of fuzzing, bounded exploration, and Markovian models further enhances coverage of potential attack vectors.

An emerging direction combines automated reasoning with machine learning to detect complex vulnerabilities and support intelligent security monitoring. AI-enhanced vulnerability detection has shown promise in identifying subtle flaws in large codebases, especially when integrated with logical verification engines. Tools that leverage formal logic alongside neural models can improve explainability and reduce verification overhead. Hybrid learning-verification pipelines, as explored in, represent a promising research trajectory. In distributed settings such as federated systems, reasoning about vulnerabilities requires logic-based analysis of confidentiality, integrity, and trust relations across multiple entities.

\section{Toward Unified Reasoning for Cybersecurity: Gaps, Challenges, and Underexplored Directions}

Despite significant advances in formal methods, automated reasoning, and cybersecurity engineering, there remains a notable fragmentation in the literature addressing their intersection. While numerous studies explore individual components—such as vulnerability detection, security protocol verification, or logic-based policy enforcement—comprehensive surveys that synthesize the role of formal logic and automated reasoning across cybersecurity domains are scarce. Existing surveys often prioritize system performance, smart contract security, or symbolic analysis, but few address the integration of logical frameworks and reasoning automation as core pillars of cybersecurity design and assurance.

A critical gap lies in the limited adoption of expressive and modular logical systems in security analysis. While propositional and first-order logic are widely used, more expressive formalisms—such as temporal epistemic logic, probabilistic deontic logic, and hybrid modal logics—remain largely underutilized, despite their potential to model knowledge, uncertainty, obligation, and time in adversarial settings. Temporal epistemic logic, for example, can naturally capture information flow, authentication handshakes, and adversary observability across protocol steps. Probabilistic deontic frameworks can encode uncertainty in system obligations (e.g., rate-limited access control), yet these logics are seldom integrated into real-world security tools.

Another underexplored area is the development of unified verification pipelines that combine diverse reasoning strategies. While tools for symbolic execution, SMT solving, model checking, and theorem proving exist in isolation, few frameworks integrate these paradigms into cohesive architectures for holistic cybersecurity assurance. Hybrid static-dynamic verification frameworks, compositional reasoning environments, and multi-layered specification languages remain topics of active research but lack industrial adoption and tooling maturity. Recent efforts to bridge this gap, such as the integration of symbolic execution engines with refinement-type systems or proof assistants, point in promising directions but require further generalization and scalability engineering.

The application of formal methods in emerging domains—such as IoT security, federated systems, and cyber-physical infrastructures—also suffers from a lack of comprehensive reasoning frameworks. Much of the current work is focused on traditional computing environments, with limited coverage of context-aware or resource-constrained systems. These domains introduce unique security challenges such as real-time constraints, decentralized trust assumptions, and adversarial control over physical interfaces. Although logic-based analysis has been extended to federated models and constrained networks, there is a need for logic-aware formal DSLs and compositional specification languages tailored to these environments.

Scalability remains a fundamental bottleneck. While symbolic reasoning tools like SMT solvers and model checkers offer automation, their application to large-scale, real-world systems is often limited by computational overhead and manual specification burdens. Verification frameworks struggle to keep up with the dynamic, heterogeneous nature of modern cyber-infrastructures. Compositional verification—using techniques such as assume-guarantee reasoning, modular contracts, and indexed logical types—offers a potential pathway forward. However, existing tools only partially support these methods, and integration into agile development pipelines remains rare.
Emerging research at the intersection of AI and formal methods offers a compelling vision for next-generation reasoning systems, but it is still in its infancy. Neural-symbolic methods for specification inference, counterexample generation, and proof synthesis show promise, yet current prototypes lack formal correctness guarantees or clear trust boundaries between learning components and deductive backends. Trust-aware pipelines, such as those based on proof-carrying code or property-carrying predictions, could help mitigate these concerns, but further theoretical development is needed.

In summary, while formal logic and automated reasoning have proven their utility in various cybersecurity subfields, several technical and conceptual gaps hinder their broader adoption and effectiveness. These include:
\begin{itemize}
    \item Underutilization of expressive logical systems for modeling time, knowledge, uncertainty, and obligations in adversarial settings.
    \item Fragmentation of reasoning tools, with limited support for hybrid and compositional verification across system layers.
    \item Lack of scalable, formally grounded frameworks for emerging domains such as IoT, federated systems, and cyber-physical security.
    \item Limited integration of AI-based reasoning with formal methods, and a need for trustworthy neural-symbolic interfaces.
    \item Insufficient support for modularity, abstraction, and automation in large-scale formal verification.
\end{itemize}

Addressing these challenges requires new research at the intersection of logic, formal methods, and systems engineering, with a focus on both theoretical expressiveness and practical toolchain development. Closing these gaps will be critical for enabling trustworthy, scalable, and explainable cybersecurity in modern digital infrastructures.

\section{Charting the Path Forward: Research Frontiers in Logic-Based Cybersecurity}

As cyber threats evolve in scale, complexity, and unpredictability, the role of logic and formal reasoning in cybersecurity must evolve accordingly. The next generation of research must bridge the gap between foundational theory and operational deployment, reimagining what it means to secure systems through verifiable guarantees. Emerging directions point toward a more expressive, automated, and compositional paradigm of cybersecurity analysis—one where logical systems not only describe security properties, but also serve as engines for scalable assurance, human-machine collaboration, and cross-domain trust.

A central priority lies in developing integrated logical frameworks that seamlessly combine temporal, deontic, epistemic, and probabilistic reasoning. Such frameworks would enable analysts to model sophisticated properties such as time-sensitive obligations, distributed knowledge under partial observability, and probabilistic guarantees on compliance or adversarial inference. This direction calls for the formalization of complex constructs like deadline-aware access control, obligation expiration, and bounded uncertainty under adversarial actions—an expressive substrate for modeling insider threats, multi-party negotiation, or dynamic coalition formation. New hybrid logics will need to support compositional semantics, proof search automation, and efficient model checking for temporal-epistemic modalities under adversarial interference.

Concurrently, the integration of symbolic logic with AI models opens up a rich space of hybrid \emph{neural-symbolic verification} architectures. These systems aim to leverage the strengths of language models and neural representation learning while maintaining the rigor and explainability of formal verification pipelines. For example, LLMs can assist in extracting or refining formal specifications from informal security policies or logs, which can then be verified using traditional SMT solvers or proof assistants. However, to ensure trust in such pipelines, future work must formalize the interface between learned outputs and deductive backends—introducing mechanisms like \emph{proof-carrying predictions}, \emph{statistically sound certificate checkers}, or \emph{learned conjecture synthesis} modules. Such tools can enable dynamic threat detection, semi-automated auditing, and continuous verification in evolving cyber environments.

As cybersecurity moves toward federated, decentralized, and cross organizational infrastructures, logic-based reasoning must expand to capture security protocols that span administrative, trust, and jurisdictional boundaries. Federated systems such as federated learning, cloud-edge architectures, and cross-chain blockchains introduce heterogeneous security policies, distributed ownership, and asynchronous trust assumptions. Formal systems capable of modeling inter-domain message-passing, privacy-preserving aggregation, and joint security properties will be essential. Future frameworks should support meta-consensus protocols, distributed epistemic semantics, and threshold-based trust negotiation—integrating symbolic verification with dynamic runtime assurance to enable privacy-aware cooperation among loosely coupled entities.

A persistent challenge in the formal verification of security-critical systems is scalability. As systems grow in size and heterogeneity, monolithic verification becomes impractical. Modular and compositional verification techniques offer a promising path forward. In these approaches, systems are decomposed into logical components, each verified with local specifications, and then reassembled using composition theorems and interface refinement. Techniques such as \emph{assume-guarantee reasoning}, \emph{logical contract refinement}, and \emph{effect-tracking type systems} can help contain verification complexity while enabling reuse of verified modules. Scalability also requires improved tool support—e.g., IDEs for modular proof development, logic-based dependency analysis, and automated abstraction-refinement loops.

Security for emerging environments such as IoT, embedded systems, and cyber-physical systems presents distinct verification challenges. These include real-time constraints, intermittent connectivity, energy limits, and decentralized trust. Logic-based models must capture time-sensitive operations, dynamic sensor data, and trust-dependent control flows. Reasoning frameworks for secure communication in these settings must incorporate real-time temporal logic, symbolic consensus protocols, and context-aware integrity verification. Compositional trust models, secure aggregation logics, and resource-bounded type systems will be necessary to make formal methods tractable for resource-constrained nodes operating in adversarial and uncertain environments.

Equally critical is the need to tightly couple formal reasoning with the development lifecycle. This entails the design of \emph{reasoning-aware domain-specific languages (DSLs)} and \emph{verified compilation chains} that allow developers to write secure software with guarantees that persist through to execution. These DSLs should support fine-grained security annotations, abstract control policies, and verification-friendly syntax. Verified compilers can enforce memory safety, access control, timing constraints, and information flow properties by construction. Building on the foundations of the K framework, Lean, and Coq, future toolchains must balance expressive power with usability—ensuring that formal guarantees are not only provable, but deployable in real-world systems.

Ultimately, the future of logic-based cybersecurity lies in the creation of formal ecosystems—comprising logics, tools, AI-enhanced verifiers, and certified runtime monitors—that collectively support end-to-end reasoning for security. Achieving this vision will require close collaboration across formal methods, programming languages, systems security, and machine learning communities. It will demand new abstractions that unify human judgment with machine inference, and verification pipelines that are robust, scalable, and context-aware. Only through such integration can we ensure that the security of tomorrow’s systems is not merely reactive, but grounded in logic, built for trust, and verifiable by design.

\section{Conclusion}

The convergence of formal logic, automated reasoning, and cybersecurity represents a critical frontier in securing modern digital infrastructure. This survey has synthesized a broad and technically diverse body of work that demonstrates how logical formalisms and automated reasoning tools provide the mathematical precision and analytical depth necessary for rigorous cybersecurity assurance. From foundational efforts in protocol verification and information flow analysis to advanced applications in vulnerability detection, federated systems, and adversarial modeling, formal methods are increasingly central to the systematic design, analysis, and certification of secure systems.

We have shown that a wide spectrum of logical frameworks—including temporal, deontic, epistemic, probabilistic, and higher-order logics—offer the expressiveness required to capture nuanced security properties such as confidentiality under time constraints, knowledge-dependent access control, and policy compliance under uncertainty. Automated reasoning techniques such as SMT solving, model checking, symbolic execution, and interactive theorem proving have enabled the formal verification of complex behaviors in software, hardware, and distributed cyber-physical systems.

Despite these advancements, significant challenges remain. The scalability of formal methods to large-scale, real-world systems is limited by the computational cost of verification and the complexity of specification authoring. Compositional verification methods offer promise but require further tooling and formalization to be practical. Moreover, the fragmented state of current reasoning tools and the underutilization of expressive logics in real-world applications hinder broader adoption. As security paradigms shift toward federated, adaptive, and AI-driven architectures, current verification strategies must evolve to accommodate dynamic trust, data mobility, and adversarial learning.

To address these gaps, we have outlined several forward-looking research directions. These include the construction of unified logical frameworks that integrate temporal, deontic, and epistemic reasoning; the development of hybrid neural-symbolic verification pipelines that leverage machine learning while preserving formal soundness; new models for cross-domain and federated security reasoning; and the advancement of compositional logic-based verification for modular, scalable system assurance. We have also emphasized the importance of reasoning-aware domain-specific languages and verified compilation pipelines to close the gap between formal specification and secure implementation.

As cybersecurity threats grow in complexity and criticality, the demand for rigorous, explainable, and composable security guarantees will only intensify. Meeting this demand requires interdisciplinary innovation that bridges formal logic, systems engineering, programming languages, and artificial intelligence. The future of cybersecurity will be shaped not just by the defenses we deploy, but by the correctness we can prove.

\bibliographystyle{plain}
\bibliography{main} % Assuming the bib file is named blockchain_logic.bib
\end{document}